\documentclass[prb,onecolumn,12pt]{revtex4-1}
\usepackage{epsfig}
\usepackage{amssymb}
\usepackage{threeparttable}
\usepackage{graphicx}
\usepackage{natbib}
\usepackage{longtable}
\usepackage{txfonts}
\usepackage{color}
\usepackage{tabu}

\begin{document}

\title{Electronic, optical and transport properties of van der Waals Transition-metal Dichalcogenides Heterostructures: A First-principle Study}

\author{Ke Xu$^1$, Yuanfeng Xu$^1$, Hao Zhang$^{1\dag}$, Bo Peng$^1$, Hezhu Shao$^2\ddag$, Gang Ni$^1$, Jing Li$^1$, Mingyuan Yao$^1$, Hongliang Lu$^3$, Heyuan Zhu$^{1\P}$ and Costas M. Soukoulis$^{4,5}$}

\affiliation{$^1$Department of Optical Science and Engineering, Key Laboratory of Micro and Nano Photonic Structures (MoE) and Key Laboratory for Information Science of Electromagnetic Waves (MoE), Fudan University, Shanghai 200433, China.  \\
$^2$Ningbo Institute of Materials Technology and Engineering, Chinese Academy of Sciences, Ningbo 315201, China\\
$^3$State Key Laboratory of ASIC and System, Institute of Advanced Nanodevices, School of Microelectronics, Fudan University, Shanghai 200433, China\\
$^4$Department of Physics and Astronomy and Ames Laboratory, Iowa State University, Ames, Iowa 50011, USA\\
$^5$Institute of Electronic Structure and Laser (IESL), FORTH, 71110 Heraklion, Crete, Greece}

\email{$^\dag$zhangh@fudan.edu.cn; $^\ddag$hzshao@nimte.ac.cn; $^\P$hyzhu@fudan.edu.cn}

\date{\today}

\begin{abstract}
Two-dimensional (2D) transition-metal dichalcogenide (TMD) MX$_2$ (M = Mo, W; X= S, Se, Te) possess unique properties and novel applications. In this work, we perform first-principles calculations on the van der Waals (vdW) stacked MX$_2$ heterostructures to investigate their electronic, optical and transport properties systematically. We perform the so-called Anderson's rule to classify the heterostructures by providing the scheme of the construction of energy band diagrams for the heterostructure consisting of two semiconductor materials. For most of the MX$_2$ heterostructures, the conduction band maximum (CBM) and valence band minimum (VBM) reside in two separate semiconductors, forming type II band structure, thus the electron-holes pairs are spatially separated. We also find strong interlayer coupling at $\Gamma$ point after forming MX$_2$ heterostructures, even leading to the indirect band gap. While the band structure near $K$ point remain as the independent monolayer. The carrier mobilities of MX$_2$ heterostructures depend on three decisive factors, elastic modulus, effective mass and deformation potential constant, which are discussed and contrasted with those of monolayer MX$_2$, respectively.
\end{abstract}

\pacs{Valid PACS appear here}
\maketitle

\section{INTRODUCTION}
The family of Two-dimensional (2D) materials has grown rapidly for their unique properties different from their 3D counterparts. A wide range of 2D materials, e.g. graphene\cite{Novo2005, zhang2005}, BN\cite{Dean2010,Yankowitz2012}, transition metal dichalcogenides (TMDs)\cite{Radisavljevic2011,Splendiani2010}, black phosphorus\cite{Xiang2015,Li2014,Tran2014}, and etc, have been proposed and under intense investigations. Among these, transition metal dichalcogenides, with the formula MX$_2$ (where M is a transition metal and X is a chalcogen), are prominent due to their finite direct band gaps, with strong optoelectronic responses\cite{Bernardi2013}, large on-off ratios and high carrier mobilities\cite{Hennig2012,Zhang2014}. Furthermore, a spin-orbit driven splitting of the valence band was found in the 2H monolayer TMDs due to the lack of inversion symmetry, which ultimately allows for valley-selective excitation of carriers\cite{Cao2012,Zeng2012,Rasmussen2015}. In addition, the electronic properties of TMDs can be tuned by strain\cite{Conley2013a}, layer numbers\cite{Mak2010}, nanostructuring\cite{Pedersen2008}, and electrostatic gating\cite{Liu2012a}, or by combining individual 2D materials into van der Waals (vdW) stacked heterostructures\cite{Novoselov2016}. The vdW heterostructures can be obtained by transfer or direct epitaxial growth\cite{Haigh2012,Hsu2014}. The interface of the heterostructures can be atomically sharp, with two-atomic thick junction region\cite{Haigh2012}, and the interlayer coupling intensity can even be tuned. Thus, the vdW heterostructures opens up many possibilities for creating new TMD material systems with rich functionalities and novel physical properties\cite{ZhangWangChenEtAl2016}. Because when two different atomically thin layers are stacked and binded by van der Waals forces to form MX$_2$ heterostructures, electronic properties of the formed vdW MX$_2$ heterostructures will be affected significantly by the alignment of the monolayer MX$_2$ to form varieties of band structures different from the monolayer counterpart, which can be direct- or indirect-bandgap, or metallic materials\cite{Terrones2013}.

For example, MoS$_{2}$-WSe$_{2}$ hetero-bilayer possesses a type II band alignment, and furthermore, the conduction band maximum                            (CBM) and valence band minimum (VBM) reside in different monolayers. Due to the separate spatial locations of CBM and VBM, the photon-generated electron-holes pairs are therefore spatially separated, resulting in much longer exciton lifetime and interlayer exciton condensation, which might help invent two-dimensional lasers, light-emitting diodes and photovoltaic devices.\cite{Rivera2015,Chiu2015}. And the evidence of strong electronic coupling between the two individual monolayer MX$_2$ in MoS$_{2}$-WSe$_{2}$ hetero-bilayer was demonstrated, leading to a new photoluminescence (PL) mode in this heterostructure\cite{Fang2014}. Hong \emph{et al} have also investigated the ultrafast charge transfer in MoS$_{2}$-WS$_{2}$ heterostructure\cite{Hong2014} and found the charge-transfer time is in femtosecond scale, much smaller than that in monolayer MoS$_{2}$ or WS$_{2}$. Furthermore, the recombination time of interlayer charge transition is tunable for different stacking order of MoS$_{2}$-WS$_{2}$  heterostructure(one was obtained by vertical epitaxial growth while the other was randomly bilayer stacked), with ~39 ps and 1.5 ns respectively\cite{Heo2015}.

To date, most researches on MX$_2$ heterostructures are concerned about the S and Se system. In this paper, by using first-principles calculations, we systematically investigate the electronic, mechanical, transport and optical properties of the vdW MX$_2$ (M = Mo, W; X= S, Se, Te) heterostructures. The bandgaps of the hetero-bilayer MX$_2$ get smaller compared with the corresponding monolayer MX$_2$. And the band alignment under Anderson's rule and interlayer coupling of heterostructures can result in direct to indirect bandgap transition. The excellent mechanical properties show the structural stability of the vdW MX$_2$ heterostructures. The transport properties exhibit encouraging results with the electron mobilites mostly higher than those of the monolayer MX$_2$. Furthermore, we also investigate the optical properties of the vdW MX$_2$ heterostructures.

\section{METHODOLOGY}
All the calculations are performed using the Vienna \textit{ab-initio} simulation package (VASP) based on density functional theory (DFT)\cite{Kresse1996}. The exchange-correlation energy is described by the generalized gradient approximation (GGA) in the Perdew-Burke-Ernzerhof (PBE) parametrization. We choose the DFT-D2 semiempirical dispersion-correction approach to involve the long-distance van der Waals (vdW) interactions\cite{Perdew1996,grimme2006}. The calculation is carried out by using the projector-augmented-wave (PAW) pseudopotential method with a plane-wave basis set with a kinetic energy cutoff of 600 eV. A 15$\times$15$\times$1 $\Gamma$-centered \textbf{k}-mesh is used during structural relaxation for the unit cell until the energy differences are converged within 10$^{-6}$ eV, with a Hellman-Feynman force convergence threshold of 10$^{-4}$ eV/\AA. The vacuum size is larger than 25 \AA\ between two adjacent atomic layers to eliminate artificial interactions between them. The electronic bandstructures of the vdW layered heterostructures are further verified by the calculations using hybrid Heyd-Scuseria-Ernzerhof (HSE06) functional\cite{HSE03,HSE06}, which improves the precision of bandstructures by reducing the localization and delocalization errors of PBE and Hartree-Fock (HF) functionals. Here the mixing ratio is 25\% for the short-range HF exchange. The screening parameter is 0.2 \AA$^{-1}$.

As we know, the electron-phonon scatterings play an important role in determining the intrinsic carrier mobility $\mu$ of 2D vdW MX$_2$ heterostructures, in which the scattering intensities by acoustic phonons are much stronger than those by optic phonons in two-dimensional materials\cite{tang2009role}. Therefore, the deformation potential theory for semiconductors, which considers only longitudinal acoustic phonon scattering process in the long-wavelength limit\cite{Cai2014, Shuai2011, Shuai2013, Wang2015b}, and was originally proposed by Bardeen and Shockley\cite{deformation1950}, can be used to calculate the intrinsic carrier mobility of 2D materials. In the long-wavelength limit, the carrier mobility of 2D semiconductors can be written as\cite{Walukiewicz1984,Takagi1996,Wang2015b}:

\begin{equation}\label{mobility1}
\mu = \frac{2e\hbar^3C}{3k_BT|m^*|^2D_l^2},
\end{equation}

where $e$ is the electron charge, $\hbar$ is the reduced Planck's constant, $T$ is the temperature equal to 300 K throughout the paper. $C$ is the elastic modulus of a uniformly deformed crystal by strains and derived from ${C}=[{\partial^2{E}/\partial^2(\Delta{l}/l_0)}]/{S_0}$, in which $E$ is the total energy,  $\Delta{l}$ represents the change of lattice constant $l_0$ along the strain direction, and $S_0$ is the lattice area at equilibrium for a 2D system. $m^*$ is the effective mass given by $m^*=\hbar^2({\partial^2{E(k)}/{\partial{k^2}})}^{-1}$ ($k$ is wave-vector, and $E(k)$ is the energy). In addition, $D_l$ is the deformation potential (DP) constant defined by $D_l^{e(h)}=\Delta{E_{CBM(VBM)}}/(\Delta{l}/l_0)$, where $\Delta{E_{CBM(VBM)}}$ is the energy shift of the band edge with respect to the vacuum level under a small dilation $\Delta{l}$ of the lattice constant $l_0$.

\section{Results and discussion}
\subsection{Geometric structures of hetero-bilayer MX$_{2}$}

\begin{figure*}
\centering
\includegraphics[width=0.7\linewidth]{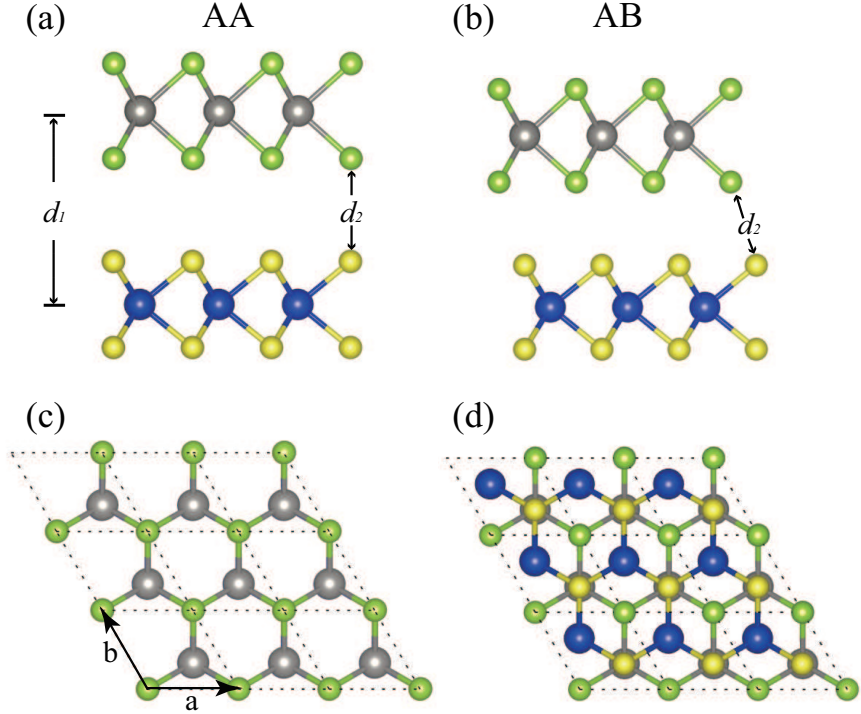}
\caption{Atomic structure of  AA stacking and AB stacking hetero-bilayer $MX_{2}$ in a 3$\times $3$\times $1 supercell from side view (upper panel) and top view (lower panel), respectively. Large and small spheres represent the M and X atoms, respectively. A color coding is used to distinguish the different atomic species. d$_{1}$ and d$_{2}$ are the interlayer distance (M$_{1}$-M$_{2}$) and the bond length of X$_{1}$-X$_{2}$.}
\label{POSCAR}
\end{figure*}

Generally, the MX$_2$ crystals have four stable lattice structures, i.e., 2H, 1T, 1T' and 3R\cite{Wilson1969}, with the first being the dominating one in nature. Most MX$_2$ crystals, like MoS$_{2}$ and WSe$_{2}$ with a stable 2H phase (1H for monolayer), have been studied widely\cite{Bhimanapati2015}. For 2H-phase MX$_2$ crystals, the M atoms and X atoms are located in different layers respectively, which can be described by the point group $D_{3h}$. While for the 3R-phase unit cell shown as Fig.~\ref{POSCAR}(b,d), one M atom is eclipsed by the X atoms above and the other one is located in the hexagonal center, leading to the $AB$ Bernal stacking. Here, we only focused on the AA and AB Bernal stacking. One stacking type can be transformed to the other one by horizontal sliding or by the rotation around the vertical axis. For MX$_2$ heterostructures with two different constituent monolayer MX$_2$ crystals, both AA and AB Bernal stacking possess a lower symmetry of $C_{3v}$ point group due to the lack of the mirror reflection $\sigma_{h}$ in the horizontal plane. The symmetry operations include $C_{3}$ and vertical mirror reflection $\sigma_{v}$\cite{BurnsG1977}. When the two constituent monolayer MX$_2$ crystals are identical, the AA stacking still possesses $D_{3h}$ symmetry.

To determine the energetically stable structure before geometry optimization, an interlayer-distance optimization step is implemented to find out an optimized $d$ (defined in the Fig.~\ref{POSCAR}(a)) using the so-called Universal Binding Energy Relation (UBER) method\cite{Rose1981,Zhao2016}. The optimized interlayer distance is predicted from a series of unrelaxed models with different $d$ (from 5 to 8 \AA), and then we calculate the surface adhesion energy W$_{ad}$ for all 30 types of 2D vdW  MX$_2$ heterostructures under investigations here (take MoS$_{2}$/WSe$_{2}$ hetero-bilayer as an example),
\begin{equation}\label{energy}
W_{ad}=\frac{E_{MoS_{2}}+E_{WSe_{2}}-E_{MoS_{2}/WSe_{2}}}{A},
\end{equation}
where A is the interface area and $E_{MoS_{2}}$, $E_{WSe_{2}}$, $E_{MoS_{2}/WSe_{2}}$ are the total energies of the monolayer MoS$_{2}$, WSe$_{2}$ and the MoS$_{2}$/WSe$_{2}$ heterostructure, respectively. The optimal interlayer distances $d$ can be obtained by maximizing the value of W$_{ad}$. Then the obtained optimized structure was further optimized again without any external constraints.

The calculated lattice constant $a$ and interlayer distance $d$ for the above-mentioned 30 types of 2D MX$_2$ heterostructures are summarized in the TABLE \ref{table-structure}, which are in good consistence with previous theoretical and experimental results of the monolayer MX$_2$\cite{Schutte1987,Coehoorn1987,Bronsema1986}. and are not sensitive to the interlayer distance. As shown in TABLE \ref{table-structure}, the optimized interlayer distances of AA stacking structures are larger than those of the corresponding AB stacking structures, which is due to the fact that, in AB structures, the X atoms are not aligned along the vertical axis and a shorter interlayer distance leads to a smaller total energy. Since the M atoms in different layers almost has no interactions, the change of stacking type will affect the interlayer interactions of X atoms.

\begin{table*}
\centering
\scriptsize
\caption{Hetero-bilayer system and band alignment type, optimized lattice constant $a$ (\AA), interlayer distance $d_{1}$ (\AA) and the atmoic distance $d_{2}$ (\AA) between the adjacent anion in different layers,  band gap of MX2 heterostructure (PBE/HSE/SOC).Other theoretical data are also listed in parentheses for comparison}
\begin{tabular}{ccccclcccccccccc}
\hline
    System (Anderson) &  Stacking type  &  $a ( \AA ) $  &  $d_{1} ( \AA ) $ & $d_{2} ( \AA ) $  &  Band type  &  $E_{g}^{PBE}/E_{g}^{HSE}/E_{g}^{SOC}$ (eV)  \\
\hline
  MoS$_{2}$-WSe$_{2}$ (II)& AA & 3.251 (3.26\cite{Lu2014b})& 6.919 & 4.896 & Direct & 0.46(0.57\cite{Lu2014b})/1.01/0.23 \\
   & AB & 3.256 & 6.270 & 3.580 & Direct & 0.57/1.12/0.34  \\
  MoS$_{2}$-WS$_{2}$ (II)& AA & 3.183 (3.19\cite{Lu2014b})& 6.758 (6.8\cite{Ji2017}) & 4.826 & Indirect & 1.29(1.16\cite{Lu2014b})/1.93/1.22  \\
   & AB & 3.187 & 6.137 (6.3\cite{Ji2017}) & 3.535 & Indirect & 1.08/1.70/1.06 \\
   WS$_{2}$-WSe$_{2}$ (II)& AA & 3.250 (3.204\cite{Terrones2013})& 6.852 & 4.846 & Direct & 0.77(1.007\cite{Terrones2013})/1.24/0.51 \\
   & AB & 3.253 & 6.232 & 3.547 & Indirect & 0.80/1.31/0.61 \\
   MoSe$_{2}$-WS$_{2}$ (II)& AA & 3.249 (3.210\cite{Terrones2013})& 6.913 & 4.893 & Direct & 1.23 (1.154\cite{Terrones2013})/1.34/0.85 \\
   & AB & 3.251 & 6.303 & 3.613 & Indirect & 0.86 /1.27/0.80 \\
   MoSe$_{2}$-WSe$_{2}$ (II)& AA & 3.320 (3.277\cite{Terrones2013})& 7.078 & 3.745 & Direct & 1.23 (1.330\cite{Terrones2013})/1.79/0.93 \\
   & AB & 3.307 & 6.485 & 3.680 & Indirect & 1.21/1.83/1.09\\
   MoS$_{2}$-MoSe$_{2}$ (II)& AA & 3.250 (3.26\cite{Lu2014b})& 6.972 & 4.940 & Direct & 0.98(0.74\cite{Lu2014b})/1.10/0.56 \\
   & AB & 3.254 & 6.350 & 3.655 & Direct & 0.65/1.09/0.56 \\
   MoTe$_{2}$-MoS$_{2}$ (II)& AA & 3.328 & 7.267 & 5.058 & -- & --/0.45/--  \\
   & AB & 3.347 & 6.575 & 3.736 & -- & --/0.47/-- \\
   MoTe$_{2}$-MoSe$_{2}$ (II)& AA & 3.413 & 7.421 & 5.177 & Indirect & 0.49/0.95/0.19 \\
   & AB & 3.413 &6.784  & 3.853 & Indirect & 0.51/0.95/0.21 \\
   MoTe$_{2}$-WS$_{2}$ (II)& AA & 3.347 & 7.170 & 4.984 & -- & --/0.43/--  \\
   & AB & 3.350 & 6.576 & 3.757 & -- & --/0.42/-- \\
   MoTe$_{2}$-WSe$_{2}$ (I)& AA & 3.425 & 7.354 & 5.136 & Indirect & 0.69/1.05/0.60  \\
   & AB & 3.423 & 6.725 & 3.811 & Indirect & 0.64/1.00/0.53  \\
   MoTe$_{2}$-WTe$_{2}$ (II)& AA & 3.538 & 7.646 & 5.348 & Direct & 0.95/1.44/0.67  \\
   & AB & 3.543 & 6.954 & 3.923 & Indirect & 0.93/1.46/0.74  \\
   WTe$_{2}$-MoS$_{2}$ (III)& AA & 3.354 & 7.204 & 5.018 & -- & --/0.46/--   \\
   & AB & 3.358 & 6.584 & 3.751 & -- &  --/0.37/--  \\
   WTe$_{2}$-MoSe$_{2}$ (II)& AA & 3.423 & 7.358 & 5.128 & Direct & 0.33/0.85/0.10\\
   & AB &  3.429& 6.740 & 3.833 & Direct & 0.35/0.84/0.11 \\
    WTe$_{2}$-WS$_{2}$ (III)& AA & 3.360 & 7.114 & 4.963 & -- & --/0.41/--  \\
   & AB & 3.365 & 6.516 & 3.717 & -- &  --/0.40/--  \\
   WTe$_{2}$-WSe$_{2}$ (I)& AA & 3.422 & 7.288 & 5.092 & Direct & 0.51/0.93/0.24  \\
   & AB & 3.447 & 6.679 & 3.781 & Direct & 0.45/0.86/0.17\\
\hline
\end{tabular}
\label{table-structure}
\end{table*}

\subsection{Electronic band structure of hetero-bilayer MX$_{2}$}

\begin{figure*}
\centering
\includegraphics[width=0.9\linewidth]{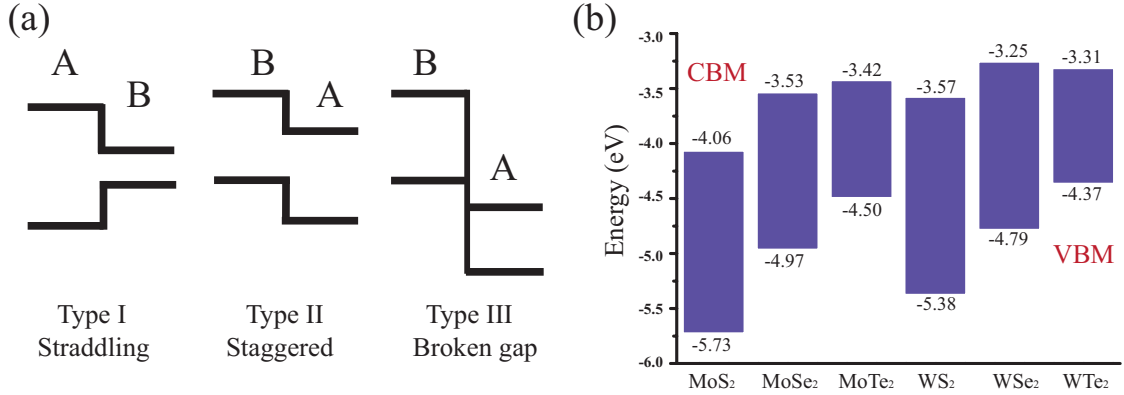}
\caption{(a) Various possible bandedge lineups in semiconductor A and B. (b) Band alignment for monolayer MX$_2$.The vacuum level is taken as 0 reference. }
\label{alignment}
\end{figure*}

Previous studies on TMDs have revealed that the monolayer $MX_{2}$ possesses direct band gap, and the conduction band maximum (CBM) and valence band minimum (VBM) located at K point\cite{Jiang2012,Kang2013,Mak2010,Ding2011}. Owing to the lack of inversion symmetry and the strong SOC effect, the valence bands possess a significant spin-orbit splitting at K valleys\cite{Kosmider2013}. And the band alignment for $MX_{2}$ shows the following trends (see from Fig.~\ref{alignment}(b)) For common-X system, the band gap of MoX$_{2}$ are larger than that of WX$_{2}$, and the CBM and VBM of WX$_{2}$ are higher than those of MoX$_{2}$; 2) For common-M system, an increase of the atomic number of X results in a shallower anion $p$ orbital and thus a shift of the VBM to higher energies, finally leading to decreased band gaps\cite{Zeier2016}. To understand these two trends in band alignment, the atomic orbital composition of the states should be taken into consideration. Taking MoS$_{2}$ as an example, the CBM of MoS$_{2}$ is mainly composed by the $d_{z^{2}}$ orbital of Mo and the $p_{x}$ and $p_{y}$ orbitals of S, whereas the VBM mostly consists of the $d_{x^{2}-y^{2}}$ and $d_{xy}$ orbitals of Mo.

For the hetero-bilayer MX$_2$ crystals constructed by two monolayer MX$_2$, their band structures can be understood by the so-called Anderson's rule, which provides the scheme of the construction of energy band diagrams for the heterostructure consisting of two semiconductor materials\cite{Anderson1960}. According to the Anderson's rule, the vacuum energy levels of the two constituent semiconductors on either side of the heterostructure should be aligned at the same energy\cite{Vol.2004}, and there are three types of possible bandedge lineups: straddling, staggered and broken gap, as shown in Fig.~\ref{alignment}(a). For type I heterostructure, the conduction band maximum (CBM) and valence band minimum (VBM) mainly consist of the orbitals of semiconductor B, which possesses a smaller band gap compared to semiconductor A. Thus, the band type of the heterostructure is consistent with the smaller-gap material. For type II heterostructure, the VBM and the CBM around the Fermi level reside in two separate semiconductors, and the formed heterostructure still possesses a small direct or indirect band gap. As for type III heterostructure, the locations of CBM and VBM are similar to those of type II heterostructure, but there does not exist band gap, and the formed heterostructure is a semimetal. It should be noted that, for type II and type III heterostructures, since the CBM and VBM may locate on different semiconductors, the photon-generated excitons are thus spatially separated, which will suppress the recombination of electron-hole pairs and extend the excitons lifetimes compared with the corresponding individual semicondutors\cite{Kang2013,Chiu2015,Fang2014a,Rivera2015,Zhang2016,Chiu2015a}.

\begin{figure*}
\centering
\includegraphics[width=0.9\linewidth]{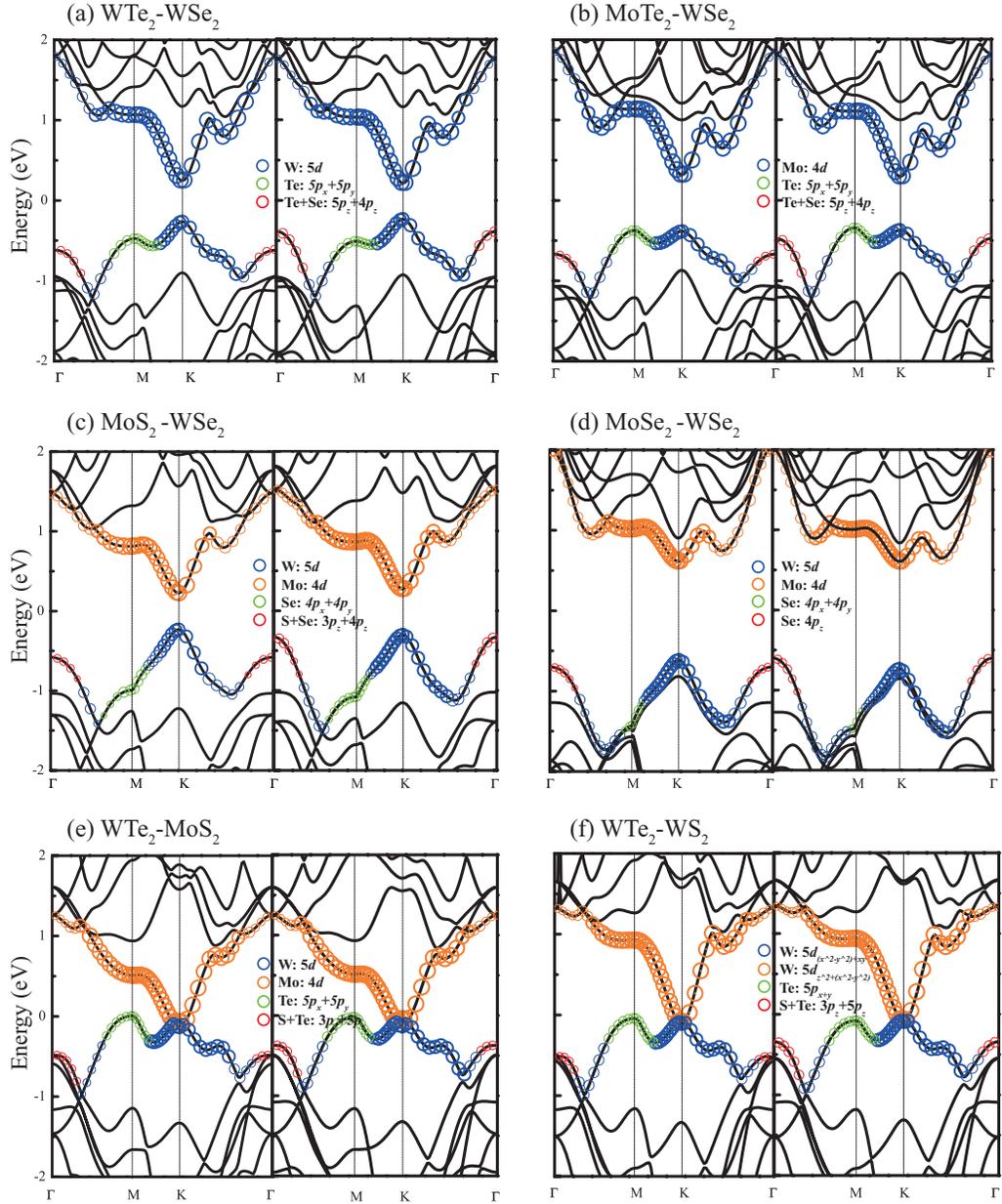}
\caption{Band structures of the AA and AB stacking vdW MX$_2$ heterostructures and atomic orbital weights in the energy bands. The blue and orange circles represent $d$ orbitals of the cations. The green and red circles represent $p_{x}+p_{y}$ and $p_{z}$ orbitals of the anions, respectively. The size of each circle is proportional to the weight of the atomic orbital. (a)(b) Type I band alignment system: WTe$_{2}$-WSe$_{2}$ and MoTe$_{2}$-WSe$_{2}$ hetero-bilayer. (c)(d)Type II band alignment system: MoS$_{2}$-WSe$_{2}$ and MoSe$_{2}$-WSe$_{2}$ hetero-bilayer. (e)(f)Type III band alignment system: WTe$_{2}$-MoS$_{2}$ and WTe$_{2}$-WS$_{2}$ hetero-bilayer.}
\label{band}
\end{figure*}

The band types and bandgaps for the vdW MX$_2$ heterostructures are calculated by the PBE and HSE06 method and the results are shown in TABLE \ref{table-structure}. The direct band gap at K point for monolayer MX$_2$ is transformed into three types of band gaps when a hetero-bilayer MX$_2$ crystal is formed, i.e., direct, indirect ($\Gamma$-K, M-K) and zero bandgap or overlapping bands, according to the calculated results shown in TABLE \ref{table-structure} and the above-mentioned analysis based on the Anderson's rule. The formation type of the band gap for the vdW MX$_2$ heterostructures categorized according to the Anderson'rule is also shown in TABLE \ref{table-structure}. The classification of the band types according to the Anderson's rule is called as Anderson band type hereafter. It is shown in TABLE \ref{table-structure} that, the Anderson band types for the vdW MX$_2$ are determined by the constituent monolayer MX$_2$ irrespective of the stacking manner, which is probably due to the fact the VBM/CBM of hetero-bilayer structure is attributed to the $d/p-$obitals of M/X atoms, and the weak vdW interactions will not change the charge distribution of the substituent monolayer MX$_2$ of the hetero-bilayer structure significantly.

For simpilicity, we first consider the Anderson band type I heterostructure, e.g. band structures for WTe$_{2}$-WSe$_{2}$ and MoTe$_{2}$-WSe$_{2}$ hetero-bilayer structures shown as Fig.~\ref{band}(a,b). Generally, as we mentioned above, two monolayer MX$_2$ crystals with identical M atoms but different X atoms possess different CBM/VBM energy levels, and the crystal with the X atoms with a larger atomic number has a higher energy level of CBM or VBM. However, as shown in Fig.~\ref{alignment}(b), the CBM energy-level of WTe$_{2}$ is lower than that of WSe$_{2}$, although the atomic number of Te is larger than Se. Such a deviation can be understood by the fact that the bond length $d_{W-Te}$ of WTe$_{2}$ is the largest one among the monolayer MX$_2$ crystals, which leads to a small overlap integral $V$ between $d$ orbitals of M atoms and $p$ orbitals of X atoms for the formation of CBM due to $V\propto1/d_{W-Te}^2$\cite{Peng2018,Froyen1979}, and thus counteracts the increase of CBM energy level from Se with a swallower $p$ orbitals compared to Te\cite{Kang2013}. The smaller CBM energy-level of WTe$_{2}$ ultimately results in the Anderson band type-I alignment of band edges in WTe$_{2}$-WSe$_{2}$ hetero-bilayer, which possesses a direct bandgap at $K$ point for both AA and AB stacking manners, as shown in Fig.~\ref{band}(a).

As shown in Fig.~\ref{band}, the valence band at the $M$ point is attributed to the $p_x$ and $p_y$ orbitals of X atoms, and the corresponding energy level for hetero-bilayer MX$_2$ crystals containing Te atoms is larger than those containing Se or S atoms, since the mass of Te is the largest one. Therefore, for hetero-bilayer MTe$_2$-MX$_2$, the valence band energies at $M$ point significantly increase compared with the hetero-bilayer MSe$_{2}$-MX$_2$ (X$\neq$Te) or MS$_{2}$-MX$_2$ (X$\neq$Te), which subsequently leads to the formation of the $M-K$ indirect band gap for the Anderson band type I heterostructure, e.g. hetero-bilayer MoTe$_{2}$-WSe$_{2}$, as shown in Fig.~\ref{band}(b).

\begin{figure*}
\centering
\includegraphics[width=0.9\linewidth]{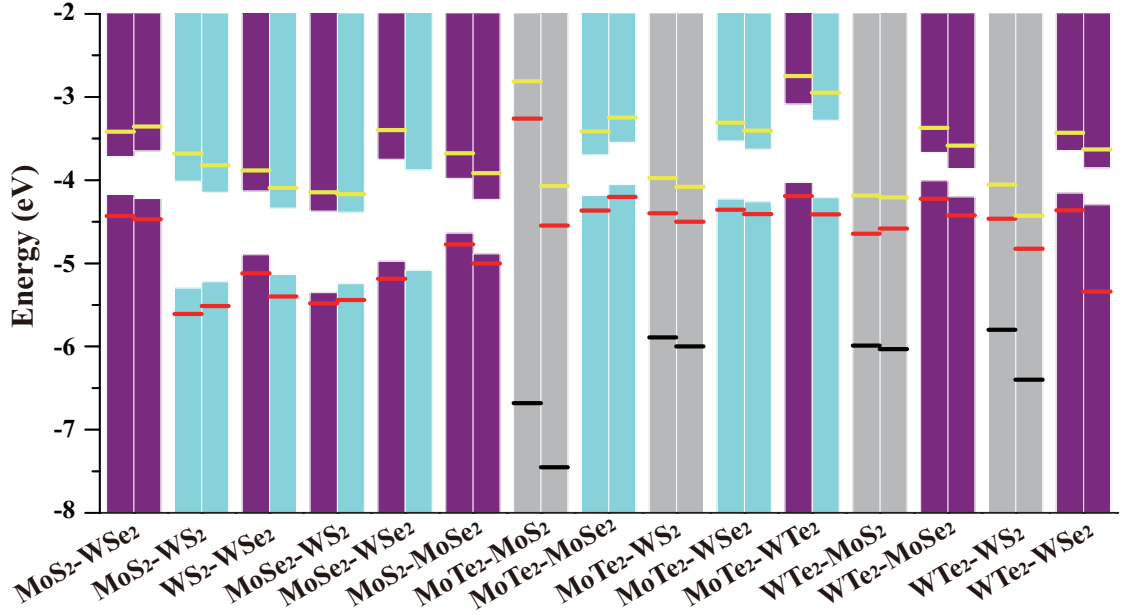}
\caption{Calculated band alignment for the vdW MX$_2$ heterostructures. The histogram is obtained by PBE, with the purple, blue and grey representing the direct bandgap, indirect bandgap and zero-bandgap, respectively. The red and yellow solid lines represent the VBM and the CBM obtained by HSE. }
\label{all}
\end{figure*}

As shown in TABLE \ref{table-structure}, most of the hetero-bilayer MX$_2$ crystals are the Anderson band type II heterostructure, e.g., hetero-bilayer MoS$_{2}$-WSe$_{2}$ and MoSe$_{2}$-WSe$_{2}$. Fig.~\ref{band}(c,d) show the energy band structures of the AA and AB stacking MoS$_{2}$-WSe$_{2}$ hetero-bilayers, exhibiting direct bandgaps of 0.46eV and 0.57eV for AA and AB stacking type, respectively, which are consistent with the previous results\cite{Lu2014}. The CBM locates on the MoS$_{2}$ layer and the VBM locates on the WSe$_{2}$ layer, resulting in the formation of spatially separated electron-hole pairs. Experiments on hetero-bilayer MoS$_{2}$-WSe$_{2}$ revealed the dramatically quenching of the photoluminescence (PL) intensities\cite{Fang2014}, and the extended exciton lifetime\cite{Chiu2015}.

The valence band at the $\Gamma$ point can be attributed to the inter-layer overlap integral of $p_z$ orbitals of X atoms belonging to different monolayers at $\Gamma$ point, as shown in Fig.~\ref{band}. For hetero-bilayer MX$_2$ considered here, the distance between X atoms belonging to different monolayers for the AB stacking hetero-bilayer, i.e. $d_2$ shown in Fig.~\ref{POSCAR}(a,b), is smaller than the corresponding AA stacking hetero-bilayer, as shown in TABLE \ref{table-structure}, thus the energy level of the valence band at the $\Gamma$ point for AB stacking hetero-bilayer is larger than that for AA stacking hetero-bilayer, due to $V_{p_z-p_z}\propto1/d_2^2$. The increase of the energy level of the valence band at $\Gamma$ points sometimes leads to the formation of $\Gamma-K$ indirect band gap, e.g. MoSe$_2$-WSe$_2$ as shown in Fig.~\ref{band}(d).

The extreme state of staggering is the formation of broken bandgaps, which is also called as the Anderson band type III alignment, as shown in Fig.~\ref{alignment}(a). For example, the CBMs of MoS$_{2}$ and WS$_{2}$ are much lower than that of other monolayer MX$_2$ and the WTe$_{2}$ possess the highest VBM, as shown in Fig.~\ref{alignment}(b), the band alignment in hetero-bilayer WTe$_{2}$-MoS$_{2}$ and WTe$_{2}$-WS$_{2}$ thus can be approximately considered as the Anderson band type III alignment, as shown in Fig.~\ref{band}(e,f). The band overlaps at $K$ point, changing the heterostructures into metallic phase.

The bandgaps of the hetero-bilayer MX$_2$ crystals based on the HSE and SOC calculations are also provided in TABLE \ref{table-structure}. The negative SOC effects decrease the band gap and the HSE calculations increase the band gap by 0.4-0.6 eV, compared to the bandgap values calculated by PBE calculations. It should be noted that the metallic phases of the hetero-bilayer MX$_2$ crystals, i.e. the Anderson band type III heterostructures, e.g. hetero-bilayer WTe$_{2}$-MoS$_{2}$ and WTe$_{2}$-WS$_{2}$ crystals as shown in Fig.~\ref{band}(e,f), are replaced by direct bandgap phases based on the more precise HSE calculations, which means that the hetero-bilayer MX$_2$ crystals considered here does not possess the Anderson band type III alignment.

\subsection{Mechanical properties and transport properties of hetero-bilayer MX$_{2}$}

\begin{table*}
\centering
\scriptsize
\caption{Hetero-bilayer system and band alignment type, Young's modulus $Y (GPa)$ and Poisson's ratio $v$ , electron and hole effective masses along armchair direction, deformation potential constants for CBM and VBM, elastic modulus, electron and hole mobilities along armchair direction.}
\begin{tabular}{ccccccccccccc}
\hline
    System (Anderson) &  Stacking type  &  $Y (N/m)$  &  $v$  & $m_{e}^{*}(m_{0})$&$m_{h}^{*}(m_{0})$& $D_l^{e}$ & $D_l^{h}$ & C ( N/m ) & $\mu_e$(cm$^2$/(V$\cdot$s)) & $\mu_h$ (cm$^2$/(V$\cdot$s))\\
\hline
  MoS$_{2}$-WSe$_{2}$ (II)& AA & 209.95 & 0.29 & 0.47 & 0.46 & 3.03 & 2.88 & 118.58	&896.07	&873.17\\
   & AB & 225.30 & 0.23 & 0.47 & 0.46 & 2.96 & 3.52& 111.47&	565.41	&873.1\\
  MoS$_{2}$-WS$_{2}$ (II)& AA  & 241.46 & 0.25  & 0.46 & 1.70 & 6.01 & 5.70& 127.81	&256.46&	18.04\\
   & AB &  242.03 & 0.24  & 0.46 & 0.92  & 6.28 & 5.03 & 121.19&	318.08&	76.7\\
   WS$_{2}$-WSe$_{2}$ (II)& AA & 206.89 & 0.25  & 0.28 & 0.46  & 3.33 & 3.27 &114.74	&1939.55&	709.71\\
   & AB & 218.60 & 0.20  & 0.28 & 0.85  & 5.65 & 4.88 &118.01&	895.83&	75.47 \\
   MoSe$_{2}$-WS$_{2}$ (II)& AA & 272.60 & 0.31  & 0.28 & 0.71  & 3.10 & 3.25 & 119.6	&2005.27	&360.99\\
   & AB & 263.53 & 0.30  & 0.28 &  0.97 & 5.28 & 4.61 &112.98	&940.06	&63.53 \\
   MoSe$_{2}$-WSe$_{2}$ (II)& AA & 206.94 & 0.25  & 0.54 & 0.44  & 2.14 & 2.66 & 109.98	&758.95	&1871.24\\
   & AB & 215.79 & 0.22  & 0.56 & 1.29  & 4.01 & 3.24  &111.32	&477.54	&61.56\\
   MoS$_{2}$-MoSe$_{2}$ (II)& AA & 232.78 & 0.26  & 0.42 & 0.71  & 2.87 & 2.78 & 125.83	&1321.55	&454.69\\
   & AB & 230.26 & 0.27  & 0.42 & 0.71  & 3.07 & 4.50  &114.86	&758.03	&359.04\\
   MoTe$_{2}$-MoS$_{2}$ (II)& AA & 196.82 & 0.36  &  &   &  & & & &\\
   & AB & 196.87 & 0.34  &  &   &  & & & & \\
   MoTe$_{2}$-MoSe$_{2}$ (II)& AA& 184.77 & 0.31  & 0.46 & 1.37  & 4.40 & 3.74 & 113.18&	532.75&	45.79\\
   & AB & 200.46 & 0.25  & 0.46 & 1.37 & 4.07 & 3.75  &110.81	&532.75&	45.79\\
   MoTe$_{2}$-WS$_{2}$ (II)& AA & 206.17 & 0.28  &  &  &  & & & & \\
   & AB & 195.86 & 0.31  &  &  &  &  & & &\\
   MoTe$_{2}$-WSe$_{2}$ (I)& AA & 183.70 & 0.28  & 0.30 & 1.33 & 3.95 & 3.83 &  109.1&	515.87	&52.52\\
   & AB & 194.71 & 0.24  & 0.30 & 1.25 & 4.41 & 4.14 &  114.79	&1191.02&	58.76\\
   MoTe$_{2}$-WTe$_{2}$ (II)& AA & 136.33 & 0.39  & 0.57 & 0.42 & 1.61 & 1.38 & 101.62&	1023.61&	55.76\\
   & AB &  171.83 & 0.22  & 0.58 & 3.46 & 4.32 & 3.30  &99.43	&2315.94	&3285.72\\
   WTe$_{2}$-MoS$_{2}$ (III)& AA & 169.33 & 0.20  &  &  &  &  & & &\\
   & AB & 189.09 & 0.28  &  &  &  &  & & &\\
   WTe$_{2}$-MoSe$_{2}$ (II)& AA & 183.83 & 0.27  & 0.45 & 0.48  & 2.65 & 2.85& 109.47	&382.87	&6.58\\
   & AB & 196.41 & 0.22  & 0.45 & 0.48 & 2.70 & 2.85 & 102.26	&912.5&	987.31\\
    WTe$_{2}$-WS$_{2}$ (III)& AA & 189.00 & 0.20  &  &  &  & & & & \\
   & AB & 233.27 & 0.29  &  &  &  &  & & &\\
   WTe$_{2}$-WSe$_{2}$ (I)& AA & 168.36 & 0.33  & 0.30 & 0.46 & 2.95 & 2.97  & 113.4	&912.5	&987.31\\
   & AB & 197.77 & 0.22  & 0.30 & 0.45 & 2.79 & 3.08 & 115.65	&875.3	&918.66\\
\hline
\end{tabular}
\label{table-properties}
\end{table*}

Since the MX$_2$ heterostructures under considerations here possess $C_{3v}$ symmetry, which means that the number of independent second-order elastic coefficients $c_{ij}$ is five and $c_{11}=c_{22}$\cite{Mouhat2014}. The calculated elastic coefficients of all MX$_2$ heterostructures are shown in TABLE S2, and all the vdW MX$_2$ heterostructures are mechanically stable, according to the Born criteria\cite{Peng2017a},
\begin{equation}\label{born}
C_{11}-C_{12}>0, C_{11}+2C_{12}>0,C_{44}>0
\end{equation}

The 2D Young's modulus of all MX$_2$ heterostructures, given by $Y^{2D}=\frac{c_{11}c_{22}-c_{12}^{2}}{c_{11}}$\cite{Andrew2012a}, are listed in TABLE \ref{table-properties}. The 2D Young's modulus for monolayer MX$_2$ crystals decrease from MS$_2$ to MSe$_2$ to MTe$_2$\cite{Zeng2015}, which is due to the fact that, the strength of $d_{xy,yz,zx}-p$-orbital coupling, which forms M-X bonding, becomes weaker with an increase of the atomic number of chalcogen\cite{Yu2017a}. The calculated 2D Young's modulus for monolayer MX$_2$ crystals are shown in TABLE S1. The contributions to the mechanical properties of MX$_2$ heterostructures can be roughly considered from constituent monolayer MX$_2$ crystals and the interlayer bonding.

The Young's modulus of the MTe$_{2}$-MX$_{2}$ heterostructures are lower than others due to the weakest $Y^{2D}$ of monolayer MTe$_{2}$ among the monolayer MX$_2$ crystals considered here. Meanwhile, the Young's modulus of the MX$_2$ heterostructures are a little lower than the sum of those of the corresponding monolayer MX$_2$ crystals, which means that the contribution from the interlayer bonding to the total Young's modulus is negative. The Poisson's ratios given by $v^{2D}=\frac{c_{12}}{c_{22}}$\cite{Andrew2012a}, which describes the lateral deformation when applying uniaxial strains, are calculated and shown in TABLE \ref{table-properties}. Generally materials with high Poisson's ratio possess good plasticity. The Poisson's ratios for the MX$_2$ heterostructures are numerically close to each other except WTe$_{2}$-MX$_{2}$, due to the lowest Poisson's ratio of 0.20 of monolayer WTe$_{2}$ crystal among the monolayer MX$_2$ crystals (see TABLE S1).

\begin{figure*}
\centering
\includegraphics[width=0.7\linewidth]{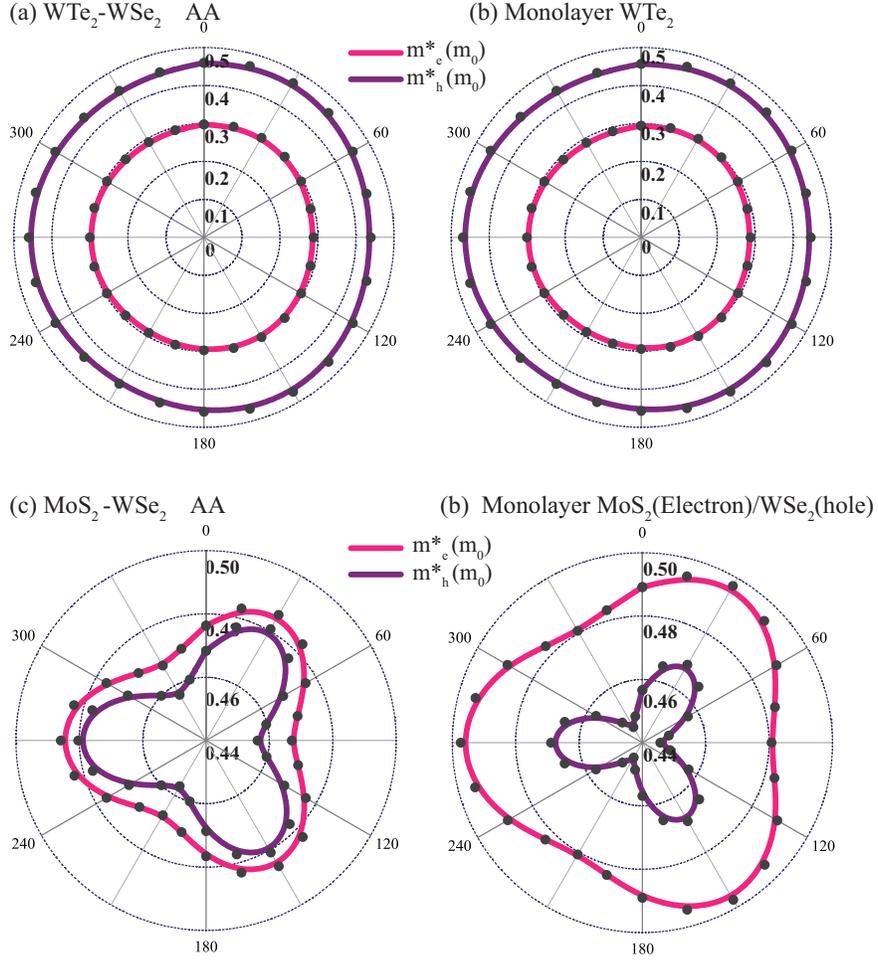}
\caption{The calculated carrier (hole mass$_{h}^{*}$ and electron mass$_{e}^{*}$) for (a) Type I band alignment system (WTe$_{2}$-WSe$_{2}$ hetero-bilayer), (b) monolayer WTe$_{2}$, (c)Type II band alignment system (MoS$_{2}$-WSe$_{2}$ hetero-bilayer), (d) monolayer MoS$_{2}$ (electron) and WSe$_{2}$ (hole)  }
\label{mass}
\end{figure*}

The calculated effective masses for electrons $m^*_e$ and holes $m^*_h$ of vdW MX$_2$ heterostructures are shown in TABLE \ref{table-structure}. The values of $m^*_e$ for AA-stacking MX$_2$ heterostructures are close to those of the corresponding AB-stacking ones, however, the values of $m^*_h$ for AA-stacking heterostructures are deviated obviously from those of AB-stacking ones, e.g. MoS$_2$-WS$_2$ and MoTe$_2$-WTe$_2$ heterostructures, especially when the band types for AA and AB stackings are different (direct vs indirect), as shown in TABLE \ref{table-structure} and \ref{table-properties}. Such phenomena can be understood by the stable location of CBM (electrons) at $K$ point for all the MX$_2$ heterostructures, and the transition of VBM (holes) from $K$ point to $M$ or $\Gamma$ point for MX$_2$ heterostructures with an indirect band gap.

As mentioned above, the bandstructures of MX$_2$ heterostructures can be roughly decomposed into those of the constituent monolayer MX$_2$ crystals, according to the Anderson's rule, which also leads to the formation of the effective masses of electrons and holes for MX$_2$ heterostructures. Fig.~\ref{mass} shows the effective masses of electrons and holes for MX$_2$ heterostructures and the corresponding constituent monolayer MX$_2$ crystals along all directions, taking WTe$_{2}$-WSe$_{2}$ and MoS$_{2}$-WSe$_{2}$ hetero-bilayer as examples without loss of generality.

The WTe$_{2}$-WSe$_{2}$ hetero-bilayer belongs to the Anderson band type I and the CBM and VBM are attributed to those of monolayer WTe$_{2}$ crystal. It is shown in Fig.~\ref{mass}(a,b) that the effective masses of electrons and holes for the WTe$_{2}$-WSe$_{2}$ hetero-bilayer are close to those of monolayer WTe$_{2}$ crystals, respectively. However, for MoS$_{2}$-WSe$_{2}$ hetero-bilayer (Anderson band type II), since the CBM is attributed to that of monolayer MoS$_{2}$ crystal and VBM is attributed to that of monolayer WSe$_{2}$ crystal, therefore, the $m^*_e$ for MoS$_{2}$-WSe$_{2}$ hetero-bilayer is similar to that of monolayer MoS$_{2}$ and the $m^*_h$ is similar to that of monolayer WSe$_{2}$, as shown in Fig.~\ref{mass}(c,d).

According to Eq. (\ref{mobility1}), the third factor determining carrier mobilites $\mu$ is the deformation potential constants, $D_l^{e,h}$, which describes the scatterings of electrons/holes by longitudinal acoustic phonons. The calculated $D_l^{e,h}$ for MX$_2$ heterostructures and monolayer MX$_2$ crystals are shown in TABLE \ref{table-properties} and TABLE S1, respectively. By comparison, it is found that, the deformation potential constants of  MX$_2$ heterostructures are overally larger than those of constituent monolayer MX$_2$, which means that, the formation of the vdW MX$_2$ heterostructures increases the electron-acoustic phonon coupling, leading to the increase of deformation potential constant $D_{l}$, especially for MoS$_2$-WS$_2$ heterostructures.

Since the CBM and VBM of the MX$_2$ heterostructures can be attributed to the respective bandstructures of the constituent monolayer MX$_2$, according to the Anderson's rule, the shift of VBM from $K$ point to $\Gamma/M$ point will result in dramatic change of the deformation potential constants and effective holes masses for MX$_2$ heterostructures with indirect bandgaps, e.g. MoSe$_2$-WSe$_2$.

\begin{figure*}
\centering
\includegraphics[width=0.8\linewidth]{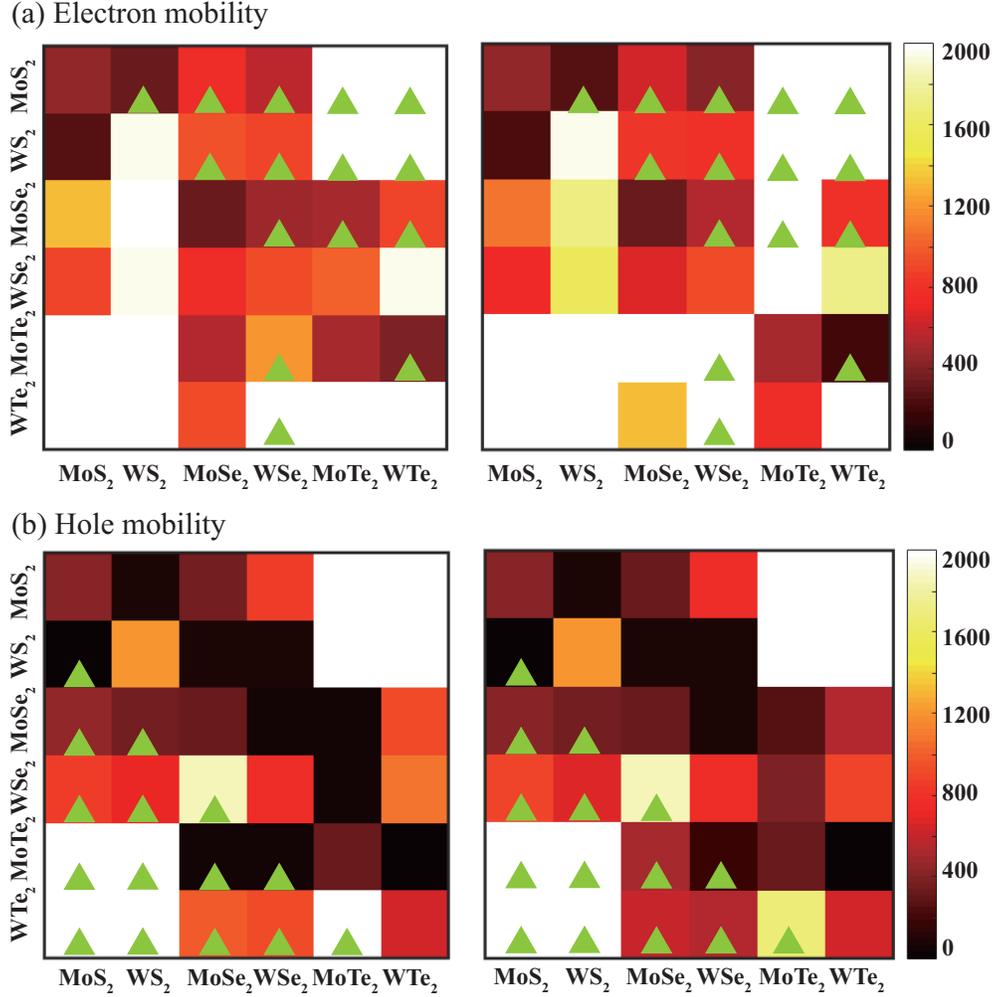}
\caption{The calculated carrier mobilities for the vdW MX$_2$ heterostructures, with the AA stacking's in lower left corner and AB stacking's upper right corner respectively. The values along diagonal are the mobilities for monolayer MX$_2$.(a)(b) are the electron mobilities of the vdW MX$_2$ heterostructures along armchair and zigzag directions, respectively; (c)(d) are the hole mobilities of the vdW MX$_2$ heterostructures along armchair and zigzag directions, respectively.}
\label{mobility}
\end{figure*}

In order to figure out the exact contributions from the three factors, i.e. effective masses $m_{e,h}^{*}$, deformation potential constants $D_l^{e,h}$ and elastic modulus $C$, to the carrier mobilities $\mu$, compared to the constituent monolayer MX$_2$ crystals, we plotted the values of the three factors for constituent monolayer crystals and hetero-bilayer structures in Fig. S4. It is clear that the elastic modulus of hetro-bilayer structures is nearly twice of the constituent monolayer MX$_2$ crystals, the deformation potential constants of hetro-bilayer structures are overally larger or close to the constituent monolayer MX$_2$ crystals except MoTe$_2$-WTe$_2$, the effective masses of hetro-bilayer structures mostly determined by the constituent monolayer cystals, are thus close to those of constituent monolayer cystals, except some hetro-bilayer structures with VBM points shifted from K to $\Gamma/M$, e.g. MoTe$_2$-WTe$_2$. Finally, the carrier mobility of electrons and holes along armchair and zigzag directions for the MX$_2$ hetero-bilayer can be calculated according to Eq. (\ref{mobility1}), as shown in Fig.~\ref{mobility}. The electron mobilities of hetro-bilayer structures are overally larger than those of constituent monolayer MX$_2$ crystals, and the same situation takes place for the holes mobilities of hetro-bilayer structures with VBM located at K point. However, the holes mobilities of hetro-bilayer structures  with VBM located at $\Gamma/M$ point are smaller than those of constituent monolayer MX$_2$ crystals.

The AA stacked MoTe$_2$-MoSe$_2$ heterostructure possesses the highest electron mobility along zigzag direction, i.e. 3658 cm$^2$/(V$\cdot$s), and the  AA stacked MoTe$_2$-WTe$_2$ heterostructure possesses the highest hole mobility along the armchair direction, i.e. 3285 cm$^2$/(V$\cdot$s).

\subsection{Optical properties of hetero-bilayer MX$_{2}$}
The optical properties of the vdW MX$_2$ heterostructures are described by the complex dielectric function, $i.e.$ $\epsilon(\omega)=\epsilon_1(\omega)+i\epsilon_2(\omega)$. The imaginary part of dielectric tensor $\epsilon_2(\omega)$ is determined by a summation over empty band states as follows \cite{Gajdos2006},
\begin{equation}
\epsilon_2(\omega) = \frac{2\pi e^2}{\Omega \epsilon_0} \sum_{k,v,c} \delta(E_k^c-E_k^v-\hbar \omega) \Bigg\vert\langle \Psi_k^c \big\vert \textbf{u}\cdot\textbf{r} \big\vert \Psi_k^v \rangle \Bigg\vert ^2,
\label{eps2}
\end{equation}
where $\Omega$ is the crystal volume, $\epsilon_0$ is the vacuum dielectric constant, $\hbar\omega$ represents the photon energy, $v$ and $c$ mean the valence and conduction bands respectively,  \textbf{u} is the polarization vector in the incident electric field, \textbf{u}$\cdot$\textbf{r} is the momentum operator, $\Psi_k$ is the wave function at the $k$ point. The real part of dielectric tensor $\epsilon_1(\omega)$ is obtained by the well-known Kramers-Kronig relation\cite{dresselhaus1999solid},
\begin{equation}
\epsilon_1(\omega)=1+\frac{2}{\pi}P\int_0^{\infty} \frac{\epsilon_2(\omega ')\omega '}{\omega '^2-\omega^2+i\eta}d\omega ',
\end{equation}
where $P$ denotes the principle value. Based on the complex dielectric function, the absorption coefficient $\alpha(\omega)$ is given by \cite{Saha2000,Luo2015}
\begin{equation}
\alpha(\omega)=\frac{\sqrt{2}\omega}{c} \Big\lbrace \big[\epsilon_1^2(\omega)+\epsilon_2^2(\omega)\big]^{1/2}-\epsilon_1(\omega) \Big\rbrace ^{\frac{1}{2}},
\end{equation}

\begin{figure*}
\centering
\includegraphics[width=1\linewidth]{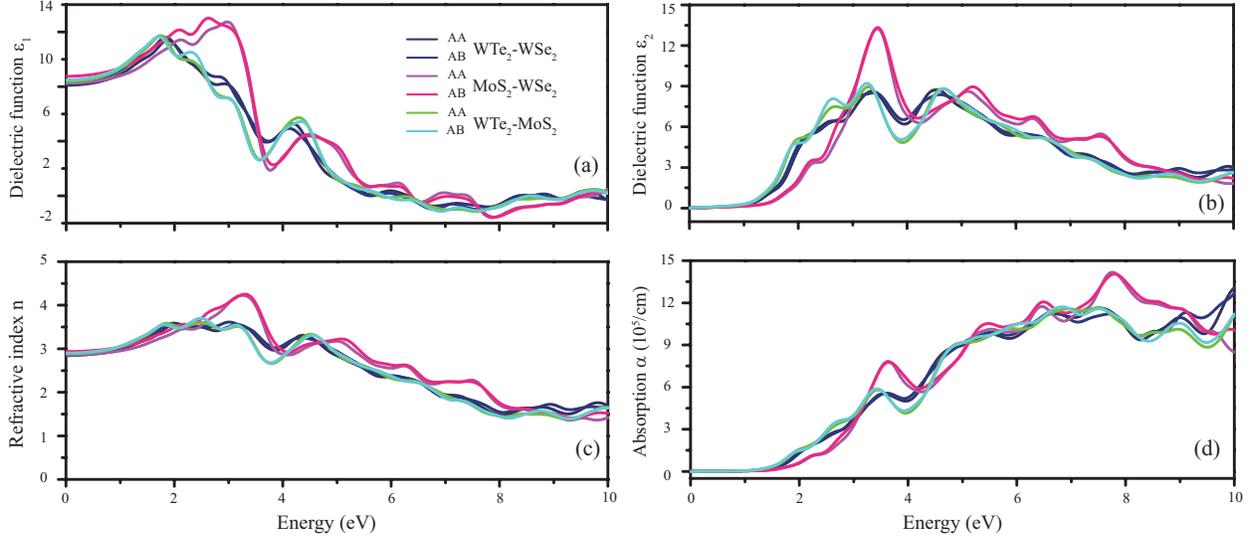}
\caption{HSE06 calculations of (a) the real part of the dielectric function, (b) the imaginary part of the dielectric function, (c) refractive and (d) optical absorption spectra of AA and AB stacking hetero-bilayer  WTe$_{2}$-WSe$_{2}$, MoS$_{2}$-WSe$_{2}$ and WTe$_{2}$-MoS$_{2}$ for incident light with the polarization along the $a$.}
\label{optical}
\label{optics}
\end{figure*}

In 2D semiconductor materials, the band gap obtained by HSE06 is usually close to the real optical band gap due to the underestimation of band gap by neglecting excitonic effects\cite{yang2016two}. Thus, we only performed HSE06 calculations to obtain optical properties for the hetero-bilayer MX$_2$ under considerations here, which show that all of them are semiconductors with a finite band gap, as shown in TABLE \ref{table-structure}. All the optical constants are calculated for incident radiations with the electric field vector \textbf{E} polarized along the $a$ and $b$ directions\cite{Xu2017b} shown in Fig.~\ref{POSCAR}(c).

Due to the C$_3$ symmetry of hexagonal structure of the hetero-bilayer $MX_{2}$, the dielectric function $\epsilon(\omega)$ possesses the same results along the $a$ and $b$ directions. And the $\epsilon(\omega)$ results for AA and AB stacking type are also close to each other, as shown in Fig.~\ref{optics}(a,b) and Fig. S4, irrrespective of the corresponding Anderson band type. The similarity in $\epsilon(\omega)$ results between AA and AB stacking hetero-bilayer $MX_{2}$ can be understood by the fact that, the bandstructure of the hetero-bilayer MX$_2$ can be roughly decomposed into the respective bandstructures of the constituent monolayer MX$_2$ according to the Anderson'rule, thus the contribution to the total optical response, i.e. $\epsilon_2(\omega)$, from absorption of an incident photon $\hbar\omega$ and then transition from $\Psi_k^c$ to $\Psi_k^v$ can be traced back to the behaviors of electrons located within the constituent monolayer MX$_2$. Therefore, the $\epsilon_2(\omega)$ results for AA and AB stacking hetero-bilayer MX$_2$ probably are similar since they contain identical constituent monolayer MX$_2$, according to Eq. \ref{eps2}.

The optical properties of hetero-bilayer MX$_2$, e.g. WTe$_{2}$-WSe$_{2}$, MoS$_{2}$-WSe$_{2}$ and WTe$_{2}$-MoS$_{2}$, are shown in Fig.~\ref{optics}. The main absorption peaks of these three hetero-bilayer MX$_2$ locate in the range of 3.0 to 5.0 eV, i.e. the ultraviolet region, with a refractive range from 2.80 to 4.27 in this region.

\section{Conclusion}
In this work, we have investigated the structure, electronic, mechanical, transport and optical properties of the vdW MX$_2$ heterostructures using first-principles calculations. The AA and AB stacked hetero-bilayer MX$_2$ exhibit three types of band alignment according to Anderson's rule, with a wide band gap range between 0 and 2 eV. The main differences between AA and AB stacked hetero-bilayer MX$_2$ lie in the band structure and mechanical properties due to the interlayer coupling such as the indirect $\Gamma-K$ bandgap. The band structure of the MTe$_2$-MX$_2$ will possesses a higher valance band at $M$ point due to the high band energy of $5p_{x,y}$ orbitals of Te. The type II band alignment of the vdW hetero-bilayer MX$_2$ make interlayer transitions possible, leading to spatially separated excitons. The transport properties of the vdW MX$_2$ heterostructures are consistent with the symmetry of the geometric structures. It should be noted that the carrier mobilities of the hetero-bilayer MX$_2$ are often higher than those of monolayer MX$_2$, attributed to the higher elastic modulus for the hetero-bilayer MX$_2$, while the hetero-bilayer MX$_2$ with indirect bandgap possess much lower hole mobilities due to the increased effective masses and deformation potential constants. Furthermore, the calculated optical properties show strong optical absorption for vdW MX$_2$ heterostructures, enabling the novel applications in optoelectronics from visible to ultraviolet region, such as photodetectors, light-emitting diodes, and photovoltaics.

\section*{Acknowledgement}
This work is supported by the National Natural Science Foundation of China under Grants No. 11374063 and 11404348, and the National Basic Research Program of China (973 Program) under Grant No. 2013CBA01505. Work at Ames Laboratory is partially supported by the U.S.Department of Energy, Office of Basic Energy Science, Division of Materials Science and Engineering (Ames Laboratory is operated for the U.S. Department of Energy by Iowa State University under Contract No. DE-AC02-07CH11358). The European Research Council under ERC Advanced Grant No. 320081 (PHOTOMETA) supports work at FORTH.

\section*{Reference}

\end{document}